\documentstyle[12pt]{article}
\begin{document}

\begin{titlepage}
\begin{flushright}
WU-AP/31/93
\end{flushright}
\begin{center}
\baselineskip .25in
\vskip 1cm
{\bf{\large Junction Conditions of Friedmann-Robertson-Walker Space-
Times}}

\vskip 0.5cm
{\sc Nobuyuki Sakai}$^{*}$ and {\sc Kei-ichi Maeda}$^{**}$

{\it Department of Physics, Waseda University, Shinjuku-ku, Tokyo 169,
JAPAN}
\end{center}

\vskip 1.5cm
\vfill
\begin{center}
{\bf Abstract}
\end{center}
\baselineskip = 24pt

We complete a classification of junctions of two Friedmann-Robertson-
Walker space-times bounded by
a spherical thin wall. Our analysis covers super-horizon bubbles and
thus complements the previous work of Berezin, Kuzumin and Tkachev.
Contrary to
sub-horizon bubbles, various topology types for super-horizon bubbles
are possible, regardless of the sign of the extrinsic curvature. We
also derive a formula for the peculiar velocity of a domain wall for
all types of junction.

\vskip 1cm
\noindent
PACS number(s): 04.20.-q

\vskip 2.5cm
\vfill
\begin{flushleft}
\baselineskip .15in
{\footnotesize  $~^{*}$~ electronic mail: sakai@cfi.waseda.ac.jp}

{\footnotesize  $~^{**}$~ electronic mail: maeda@cfi.waseda.ac.jp}
\end{flushleft}
\end{titlepage}

\baselineskip = 24pt

Using the thin-wall formalism first devised by Israel \cite{i}, many
authors studied bubble dynamics \cite{bkt}-\cite{smptp}. In the
formalism, a (2+1)
dimensional hypersurface of a trajectory of the bubble wall is
embedded in two
different homogeneous space-times: One describes the inside of the
bubble and
the other does the outside. Two space-times are pasted on the
hypersurface with
the junction conditions for metrics and extrinsic curvature tensors.

Berezin, Kuzumin and Tkachev \cite{bkt} studied the junction conditions
of two
different Friedmann-Robertson-Walker (FRW) space-times bounded by a
spherical thin wall. They considered only the case when bubbles are
smaller than the
horizon scale. From this restriction, they discussed the constraints
on decaying-vacuum parameters, which are related to the size of a
nucleated bubble,
in an open or flat universe. Although Berezin {\it et al.} mentioned
the possibility of bubbles which are larger than the horizon scale,
they did not
show the junction conditions explicitly. As long as the self-gravity
of a bubble
is ignored, a bubble is in fact nucleated within the horizon scale.
However, in
some inflationary models there may exist bubbles whose self-gravity is
dominant \cite{gz} and, therefore, whose size could be larger than the
horizon scale
\cite{smprd}. Further, vacuum parameters should be determined from
microscopic
physics but not from a macroscopic constraint such as the condition
that a
bubble should be smaller than the horizon size. In this paper, we
re-examine
junctions of FRW space-times and classify all possible spherical
bubbles. Our
study is a complement to that of Berezin {\it et al.}, and will also
be useful for analyzing super-horizon voids \cite{sv}.

To begin, we define a unit space-like vector $N^{\mu}$ which is
orthogonal to
the world hypersurface $\Sigma$ denoting a trajectory of a bubble wall
and which points from $V^-$ (inside of the bubble) to $V^+$ (outside).
For convenience, we introduce a Gaussian normal coordinate system $(n,
\tau,\theta,\varphi)$, where
$n=0$ corresponds to $\Sigma$ and $\tau$ is the proper time of the
wall. Then
the angular component of the extrinsic curvature tensor is written as
\begin{equation}
{K^{\theta}_{\theta}} \equiv N ^{\theta }_{;\theta} = - \Gamma
^{\theta} _{n \theta} = { 1 \over R} {\partial \bar R \over
\partial n}\bigg|_{\Sigma}, \label{ktrn}
\end{equation}
where $\bar R(n,\tau)$ is the circumference radius of the circle of
$n=$ constant, $\tau=$ constant, $\theta=\pi/2$ in the Gaussian normal
coordinates, and $R(0,\tau)=\bar R(\tau)$. As some authors have
pointed out \cite{bkt}-\cite{bgg}, the sign of $K^{\theta}_{\theta}$
is a key to classify the global space-time structures, because
(\ref{ktrn}) shows
that its sign depends on whether or not the spherical area increases
in the
normal direction and the sign determines the global structure for
static space-times. For example, space-time structures are classified
by the signs of ${K^{\theta}_{\theta}}^{\pm}$ for junctions of
Schwarzschild space-times or de Sitter space-times. In those cases,
the signs of ${K^{\theta}_{\theta}}^{\pm}$ determine spatial topology.
However, if either of two space-times is time-dependent, the relation
between the signs of ${K^{\theta}_{\theta}}^{\pm}$
and the spatial topology is not so clear.

First, we shall investigate whether the signs of ${K^{\theta}_{
\theta}}^{\pm}$ fix the spatial
topology in FRW space-time without using the Einstein equations. The
FRW metric is written as
\begin{equation}
ds^2 = -dt^2 +a^2(t) [ d\chi^2+r^2(\chi) (d\theta^2+\sin^2\theta d
\varphi^2) ], \label{frw}
\end{equation}
where
\begin{eqnarray}
r ( \chi ) = \left\{ \begin{array}{lll}
\sin \chi & (k=+1,& \rm closed ~universe) \\
\chi & (k=0,& \rm flat ~universe) \\
\sinh \chi & (k=-1,& \rm open ~ universe).
\end{array} \right. \label{rchi}
\end{eqnarray}
In these coordinates (\ref{ktrn}) is written down as
\begin{equation}
{K^{\theta}_{\theta}} = { 1 \over R}{\partial \bar R \over
\partial n} \bigg|_{\Sigma}
= {1\over R}\biggl({\partial \bar R \over \partial \chi}
{\partial\chi\over \partial n} +{\partial\bar R \over\partial t}
{\partial t \over \partial n}\biggr) \bigg|_{\Sigma}. \label{ktchit}
\end{equation}
By giving $\partial t / \partial n $ and $ \partial\chi /
\partial n $ in terms of physical
variables measured in $V^{\pm}$, we finally get \cite{smprd,smptp}
\begin{equation}
{K^{\theta}_{\theta}} = \zeta{\gamma\over R} \Bigl({dr\over d
\chi} + vHR\Bigr), \label{ktfrw2}
\end{equation}
where $H\equiv (da/dt)/a$ is the Hubble parameter, $ v \equiv a(d
\chi/dt)$ is the peculiar velocity relative to the background expansion,
$\gamma\equiv\partial t/ \partial \tau = 1/\sqrt{1-v^2}$ is the
Lorentz factor, and $\zeta\equiv$ sign $(\partial\chi/
\partial n)$. We have assumed $\partial
t/\partial\tau>0$ because both $t$ and $\tau$ are future directed.

 From formula (\ref{ktfrw2}), we find that the signs of $K^{\theta}_{
\theta}$ and $\zeta$ are
independent. The reason is as follows: The topology of the space-time
is determined by $\zeta^{\pm}$
and $k^{\pm}$ as listed in Fig.1. On the other hand, we can set up any
sign of ${K^{\theta}_{\theta}}$
for any value of $\zeta^{\pm}$ and $k^{\pm}$ as initial data, because
the radius $R$ and the velocity
$v$ are independent variables. Hence, all types of topology in Fig.1
are geometrically possible
regardless of the signs of ${K^{\theta}_{\theta}}^{\pm}$. The signs of
${K^{\theta}_{\theta}}^{\pm}$
does not determine the spatial topology in general.

Next, by use of the Einstein equations we shall show the relation
between matter energy density and
the extrinsic curvature $K^{\theta}_{\theta}$. The Einstein equations
give the background expansion rate,
\begin{equation}
H^2_{\pm}+{k\over a^2}\bigg|^{\pm} ={8\pi G\rho^{\pm}\over3},
\label{ein}\end{equation}
and the junction condition at the wall \cite{bkt}-\cite{bgg},
\begin{equation}
{K^{\theta}_{\theta}}^+ - {K^{\theta}_{\theta}}^- = - 4\pi G \sigma,
\label{jc}\end{equation}
where $\rho^{\pm}$ is energy density in $V^{\pm}$ and $\sigma$ is the
surface energy density of the
wall. From (\ref{ktfrw2}) with (\ref{rchi}), (\ref{ein}) and
\begin{equation}
{dR\over d\tau} = \gamma\Bigl({dr\over d\chi}v+HR\Bigl)\bigg|^{\pm},
\label{rtau}\end{equation}
we obtain
\begin{equation}
({K^{\theta}_{\theta}}^{\pm})^2 = {1\over R^2} \biggl\{1+ \Bigl({dR
\over d\tau}\Bigr) ^2 - {8\pi G\rho^{\pm}\over3} R^2\biggr\},
\label{ktfrw1}\end{equation}
which coincides with the conventional expression of $K^{\theta}_{
\theta}$ \cite{bkt}. Using (\ref{jc}) and (\ref{ktfrw1}), we have
\begin{equation}
{K^{\theta}_{\theta}}^{\pm} \equiv {({K^{\theta}_{\theta}}^+)^2 -
({K^{\theta}_{\theta}}^-)^2 \pm ({K^{\theta}_{\theta}}^+ - {K^{\theta}_{
\theta}}^-)^2 \over 2
({K^{\theta}_{\theta}}^+ -{K^{\theta}_{\theta}}^-)}
 = { \rho^+-\rho^- \mp 6\pi G\sigma^2 \over 3\sigma}, \label{kte}
\end{equation}
where the signs $\pm$ and $\mp$ correspond to values of
${K^{\theta}_{\theta}}^{\pm}$, respectively. If we assume $\sigma$ to
be positive, the signs of ${K^{\theta}_{\theta}}^{\pm}$ are determined
by the ratio of $\rho^+-\rho^-$ to $\sigma^2$.

We have summarized the classification of junctions of FRW space-times
in Table 1.
First, we find that the signs of ${K^{\theta}_{\theta}}^{\pm}$ are
determined from the value of $(\rho^+-\rho^-)/ \sigma^2$. The same
relation has already been presented by Sato \cite{hs} and by Berezin {
\it et al.} \cite{bgg},
although Sato considered only closed de Sitter space-times and Berezin
{\it et al.} restricted their study to sub-horizon bubbles. Here we
have extended the
relation to the most general case, which covers any matter fluid, any
spatial curvature, and any bubble size.

In the last column of Table 1, we present all possible junctions. To
compare our
results with the previous results for sub-horizon bubbles, we have
listed only
types with $\zeta^{\pm}=+1$ in Table 1. This is the case if we
consider realistic bubble nucleation in the expanding universe (see
Ref.[22] in Ref.\cite{smptp}). The topology types which do not satisfy
$\zeta^{\pm}=+1$ are
with parentheses in Fig.1. For a static space-time, only the types
written in
italic characters in Table 1 are possible. New types, which appear in
an expanding universe, are described by bold characters. An asterisk
has been used
for the types for which only super-horizon bubbles are possible. We
should remark
that Type II and III solutions are also possible even in a flat or
open universe, contrary to the previous results for sub-horizon
bubbles. The reason
$\partial\bar R/\partial n$ can be negative is simple. If the comoving
radius of
a bubble decreases in time, the normal vector $N^{\mu}$ points in the
``past'' direction in terms of the cosmic time $ t^+ $ in $V^+$ ({\it
i.e.}, $ N^t<0$), and accordingly $ \partial t/\partial n $ is
negative. When the second term in (\ref{ktchit}), which is negative,
becomes large, $ \partial \bar R/\partial n $
could be negative. This condition needs that the size of the bubble
must be larger than the horizon scale, because $K^{\theta}_{\theta}<0$
in (\ref{ktfrw2}) implies $v<0$ and $HR>(dr/d\chi)/|v|>1$ for $k\le0$.

Finally, we shall investigate the dynamics of vacuum bubbles in FRW
coordinates. Now, let us assume $ \rho^{\pm} $ and $ \sigma $ to be
constant. Berezin {\it et
al.} studied it only for Type I in Table 1, while our analysis below
covers all
types. From (\ref{jc}) with (\ref{ktfrw1}), we obtain the solution of
a bubble
motion \cite{bkt}:
\begin{equation}
R(\tau) = R_0\cosh{\tau\over R_0} ~~ {\rm with} ~~ R_0 = {3\sigma
\over\sqrt{(\rho^++\rho^-+6\pi G\sigma^2)^2 - 4\rho^+\rho^-}},
\label{sol}\end{equation}
This formula is valid for any spatial curvature $k^{\pm}$, as long as
a thin-wall approximation is
valid. One may expect that $dR/d\tau$ may vanish at nucleation time.
This is true
for the case of Type I, however, we cannot set $dR/d\tau=0$ for
bubbles larger
than the horizon scale in the case of Type II and III. We may
understand this to
be either that thin-wall approximation is no longer valid at nucleation
or that the wall is expanding at nucleation time ($dR/d\tau>0$).

Now we derive the peculiar velocity of the domain wall observed in
$V^{\pm}$. From (\ref{ktfrw2}) and (\ref{rtau}), we obtain
\begin{equation}
v^{\pm} = {K^{\theta}_{\theta}HR-\zeta(dr/d\chi)[(dR/d\tau)/R]\over
\zeta H(dR/d\tau)-(dr/d\chi)K^{\theta}_{\theta}}\bigg|^{\pm}. \label{v}
\end{equation}
This relation is applied to any equation of state for matter. For the
case when
$\rho^{\pm}$ and $\sigma$ are constant, using (\ref{sol}), we find the
asymptotic value of the peculiar velocity $v^{\pm}$ in the limit of $
\tau\rightarrow \infty$,
\begin{equation}
v^{\pm}_{\infty}\equiv\lim_{\tau\rightarrow \infty }v^{\pm}
 = \zeta^{\pm}{K^{\theta}_{\theta}}^{\pm}R_0
= \zeta{\rho^+-\rho^- \mp 6\pi G\sigma^2 \over\sqrt{(\rho^++\rho^-+6
\pi G\sigma^2)^2-4\rho^+\rho^-}}.
\end{equation}
If we consider bubbles with $\zeta^{\pm}=+1$, the signs of $v^{\pm}_{
\infty}$ coincide with the signs
of ${K^{\theta}_{\theta}}^{\pm}$. Then $ v^+_{\infty} $ is always
negative for new types of junction
(Type II, III). Even if $v^+_{\infty}$ is negative, however, the
physical size of such bubbles is still increasing in time.

To summarize, we re-examined the junctions of two FRW space-times and
completed the classification. For vacuum bubbles, we presented the
formula of the peculiar
velocity in all range of parameters. One of the remarkable results is
that false vacuum bubbles ($\rho^+<\rho^- $) can exist even in a
homogeneous and flat universe, if the size of a bubble is larger than
the horizon scale.

\vskip 0.5cm
This work was supported partially by the Grants-in-Aid for Scientific
Research Fund of the Ministry of Education, Science and Culture
(No.06302021 and No.06640412), and by a Waseda University Grant for
Special Research Projects.

\newpage
\baselineskip .25in
\begin{center}
{\large TABLE}
\vskip 0.7cm

\begin{tabular}{|c|c||c|c||c|c|}
\hline \hline &&&&&\\[-.5em] Type & energy condition & ${K^{\theta}_{
\theta}}^+$ &
${K^{\theta}_{\theta}}^-$ & $k^-$ &
\begin{tabular}{p{2cm}|p{1cm}|p{1cm}}
\multicolumn{3}{c}{$k^+$}\\ \hline
$+1$ & $0$ & $-1$
\end{tabular}
\\[.5em]
\hline \hline
I & {\Large${\rho^+-\rho^- \over 6\pi G\sigma^2}>1$} & $+$ & $+$
&
\begin{tabular}{c}
$+1$ \\\hline $0$ \\\hline $-1$
\end{tabular}
&
\begin{tabular}{p{2cm}|p{1cm}|p{1cm}}
$a, ~{\bf b}, ~{\bf c}, ~{\bf d}$ & $i, ~{\bf j}$ & $m, ~{\bf n}$ \\
\hline
$e, ~{\bf f}$ & $k$ & $o$ \\ \hline
$g, ~{\bf h}$ & $l$ & $p$ \\
\end{tabular}
\\[.5em]
\hline
II & {\Large${|\rho^+-\rho^-| \over 6\pi G\sigma^2}<1$} & $-$ & $+$
&
\begin{tabular}{c}
$+1$ \\\hline $0$ \\\hline $-1$
\end{tabular}
&
\begin{tabular}{p{2cm}|p{1cm}|p{1cm}}
${\bf a}, ~b, ~{\bf c}, ~{\bf d}$ & ${\bf i}^*, ~{\bf j}^*$ & ${\bf
m}^*, ~{\bf n}^*$ \\ \hline
${\bf e}, ~f$ & ${\bf k}^*$ & ${\bf o}^*$ \\ \hline
${\bf g}, ~h$ & ${\bf l}^*$ & ${\bf p}^*$ \\
\end{tabular}
\\[.5em]
\hline
III & {\Large${\rho^+-\rho^- \over 6\pi G\sigma^2}<-1$} & $-$ & $-$
&
\begin{tabular}{c}
$+1$ \\\hline $0$ \\\hline $-1$
\end{tabular}
&
\begin{tabular}{p{2cm}|p{1cm}|p{1cm}}
${\bf a}, ~{\bf b}, ~c, ~{\bf d}$ & ${\bf i}^*, ~{\bf j}^*$ & ${\bf
m}^*, ~{\bf n}^*$ \\ \hline
${\bf e}, ~{\bf f}$ & ${\bf k}^*$ & ${\bf o}^*$ \\ \hline
${\bf g}, ~{\bf h}$ & ${\bf l}^*$ & ${\bf p}^*$ \\
\end{tabular}
\\[.5em]
\hline \hline
\end{tabular}

\end{center}

\vskip 0.5 cm
\noindent
Table 1. Classification of junctions of FRW space-times. The signs of
${K^{\theta}_{\theta}}^{\pm}$
are determined from the value of $( \rho^+-\rho^- )/ \sigma^2 $,
regardless  of matter's equation of
state and spatial curvature. In the last column, we also present all
possible junctions with
$\zeta^{\pm}=+1$ (see Fig.1 for each letter). If a space-time is
static, only the types written
in italic characters are possible. New types are described by bold
characters. We
use an asterisk for the types for which only super-horizon bubbles are
possible. We find that Type II and III solutions are also possible
even in a flat or open
universe, though they are impossible for sub-horizon bubbles.

\vskip 1cm
\begin{center}  {\large FIGURE CAPTIONS}
\end{center}

\baselineskip = 24pt

\noindent Fig.1. A list of topology types of spatial sections ($t_{
\pm}=$ const.) for all possible
junctions. The types in parentheses do not satisfy $\zeta^{\pm}=+1$
and are not possible as long as the universe itself does not change
its topology when a bubble is created.

\end{document}